\documentclass[aps,pre,reprint,superscriptaddress,showpacs,amsmath
,floatfix
]{revtex4-2}

\usepackage{graphicx, float, subcaption}
\usepackage{amsmath, amsthm, amssymb, amsfonts}
\usepackage{algorithm2e}
\usepackage{algpseudocode}
\renewcommand{\vec}[1]{\boldsymbol{#1}}

\newcommand{\mat}[1]{\textbf{#1}}

\newcommand{\eq}[1]{Eq.~(\ref{#1})}

\newcommand{\eqa}[1]{equation~(\ref{#1})}
\newcommand{\Order}[1]{O(#1)}
\newcommand{\Opt}[1]{\hat{#1}}

\begin{document}


\title{A short-time drift propagator approach to the Fokker-Planck equation}



\author{Wisit Mangthas}

\author{Waipot Ngamsaad}
\email{waipot.ng@up.ac.th}
\affiliation{Division of Physics, School of Science, University of Phayao, Phayao 56000, Thailand}


\date{\today}

\begin{abstract}
The Fokker-Planck equation is a partial differential equation that describes the evolution of a probability distribution over time. It is used to model a wide range of physical and biological phenomena, such as diffusion, chemical reactions, and population dynamics. Solving the Fokker-Planck equation is a difficult task, as it involves solving a system of coupled nonlinear partial differential equations. In general, analytical solutions are not available and numerical methods must be used. In this research, we propose a novel approach to the solution of the Fokker-Planck equation in a short time interval. The numerical solution to the equation can be obtained iteratively using a new technique based on the short-time drift propagator. This new approach is different from the traditional methods, as the state-dependent drift function has been removed from the multivariate Gaussian integral component and is instead presented as a state-shifted element. We evaluated our technique employing a fundamental Wiener process with constant drift components in both one- and two-dimensional space. The results of the numerical calculation were found to be consistent with the exact solution. The proposed approach offers a promising new direction for research in this area.

\end{abstract}


\maketitle


\section{Introduction}
The Fokker-Planck equation represents the evolution of probability density function (PDF) for stochastic processes, which is important in many domains such as physics, chemistry, and biology \cite{risken1996fokker, frank2005nonlinear}. Analytical solutions to the Fokker-Planck equation can be found in some special cases but are usually difficult to obtain.  Numerical solutions to the Fokker-Planck equation become an attractive choice. Standard methods to solve the Fokker-Planck equation generally use discretized grids that evaluate PDF in state space \cite{challa2000nonlinear}. However, as the dimension increases, the amount of computation grows exponentially. For large drifting systems, traditional grid-based methods provide invalid solutions, where probability values can be negative and exhibit oscillatory behavior. Various schemes have been proposed to solve this problem, but no method has been found to be perfect \cite{xu2009quadrature, moore2022adaptive}.

The fundamental solution of the Fokker-Planck equation can be represented by Feynman Path Integration (PI), which uses a short-time propagator to provide an exact solution. Wehner and Wolfer introduced the numerical PI formula using a histogram-based PDF approximation on a grid \cite{wehner1983numerical}. In conventional fixed-grid PI implementations, short-time propagators are implemented as a propagator matrix that contains transition probabilities between all the source and target states. However, traditional grid-based PI implementations struggle with the limitations of fixed grid formulations and, as a result, have not found widespread practical application. Recently, Subramaniam and Vedula proposed a transformed path integral (TPI) method to address this problem \cite{subramaniam2017transformed}. However, the TPI is too challenging to put into practice. 

 In this paper, we propose a novel approach to solving the Fokker-Planck equation in a short time interval. The numerical solution to the Fokker-Planck equation can be obtained iteratively using a new technique based on a short-time drift propagator. We believe that it will be of great interest to researchers in many fields.

\section{Problem statement}
We consider the following continuous nonlinear dynamical system:
\begin{equation}\label{eq:dynamics}
    \dot{\vec{x}}(t) = \vec{v}(\vec{x}(t),t) + \mat{G}(\vec{x}(t),t)\vec{\eta}(t),
\end{equation}
where $\vec{x}(t) \in \mathbb{R}^{n\times 1}$ represents the state vector of the system at any given time $t$, $\vec{v}(\vec{x}(t),t) \in \mathbb{R}^{n\times 1}$ is a vector-valued function with real components, $\mat{G}(\vec{x}(t),t) \in \mathbb{R}^{n\times r}$ is a real matrix, and $\vec{\eta}(t) \in \mathbb{R}^{r\times 1}$ is zero mean white Gaussian noise that has autocorrelation $E[\vec{\eta}(t)\vec{\eta}^\top(s)] = \mat{Q}\delta(t-s)$. Assuming that the PDF for the system given in \eq{eq:dynamics} exists, it can be shown that the PDF $p \equiv p(\vec{x},t)$ at time $t$ satisfies the Fokker-Planck equation \cite{risken1996fokker, frank2005nonlinear}
\begin{equation}
\label{eq:FP}
\frac{\partial p}{\partial t} = -\sum_{i}^{n} \frac{\partial (v_i p)}{\partial x_i} + \sum_{i}^{n} \sum_{j}^{n} \frac{\partial^2 (D_{ij} p)}{\partial x_i\partial x_j},
\end{equation}
where $\mat{D} = \frac{1}{2}\mat{G}\mat{Q}\mat{G}^\top$ is a diffusion coefficient matrix. 

The solution of the Fokker-Planck equation as in \eq{eq:FP} is represented in this form \cite{risken1996fokker, frank2005nonlinear}
\begin{equation}
\label{eq:FP_int_sol}
p(\vec{x},t) = \int P(\vec{x},t|\vec{x}^\prime,t^\prime)p(\vec{x}^\prime,t^\prime) d\vec{x}^\prime,
\end{equation}
where $P(\vec{x},t|\vec{x}^\prime,t^\prime)$ is the probabilistic transition kernel \cite{risken1996fokker, frank2005nonlinear}. For small time differences $\tau=t-t^\prime \ll 1$, the short-time propagator is known
\begin{equation}
\label{eq:propagator_old}
P(\vec{x},t+\tau|\vec{x}^\prime,t) = \frac{ e^{-\frac{1}{4\tau} (\vec{x}-\vec{x}^\prime-\vec{v}\tau)^\top \mat{D}^{-1} (\vec{x}-\vec{x}^\prime-\vec{v}\tau)} }{\sqrt{(4\pi\tau)^N |\mat{D}|}} .
\end{equation}

The complexity of the integral term due to the dependence of the state vector $\vec{x}$ on the valued function $\vec{v}$ makes it difficult to calculate, making solution \eq{eq:FP_int_sol} difficult to handle with the short-time propagator in its current form.

\section{Methods and Solutions}
We introduce a novel approach for solving the Fokker-Planck equation by taking advantage of a short-time drift propagator.

\subsection{The modified Fokker-Planck equation}
We modify the short-time propagator to make it more straightforward. We may write
\begin{equation}
\label{eq:propagator_new}
P(\vec{x}+\vec{v}\tau,t+\tau|\vec{x}^\prime,t) = \int \delta(\vec{\chi}- \vec{x} -\vec{v}\tau) P(\vec{\chi},t+\tau|\vec{x}^\prime,t) d\vec{\chi},
\end{equation}
where $\delta(\vec{x})$ is the multivariate delta function. Therefore, the system transits from $\vec{x}^\prime$ to $\vec{x}+\vec{v}\tau$ during a short interval time $\tau=t-t^\prime \ll 1$. The multivariate delta function in \eq{eq:propagator_new} can be written in the Taylor expansion as follows
\begin{eqnarray}
\label{eq:delta_Taylor}
\lefteqn{
\delta(\vec{\chi}- \vec{x} -\vec{v}\tau)  } \nonumber\\
&=& \delta(\vec{x}^\prime - \vec{x} + \vec{\chi} -\vec{x}^\prime -\vec{v}\tau) \nonumber\\
&\approx&  \left(1 + \Delta_i \frac{\partial}{\partial x_{i}^\prime}  +\frac{1}{2} \Delta_i \Delta_j \frac{\partial^2 }{\partial x_{i}^\prime \partial x_{j}^\prime }  + \Order{3} \right) \delta(\vec{x}^\prime - \vec{x}),  \nonumber\\
\end{eqnarray}
where $\Delta_i = \chi_{i}-x_{i}^\prime-v_{i}\tau$. Substituting \eq{eq:delta_Taylor} into \eq{eq:propagator_new}, we have
\begin{eqnarray}
\label{eq:FP_KM}
\lefteqn{
P(\vec{x}+\vec{v}\tau,t+\tau|\vec{x}^\prime,t)  } \nonumber\\
&=&  \left(1 + M_i^1 \frac{\partial}{\partial x_{i}^\prime }  + \frac{1}{2} M_{ij}^2 \frac{\partial^2 }{\partial x_{i}^\prime \partial x_{j}^\prime }  \right) \delta(\vec{x}^\prime - \vec{x}),
\end{eqnarray}
where $M_i^1 = M_i^1(x_i^\prime,t,\tau) = \int \Delta_i P(\vec{\chi},t+\tau|\vec{x}^\prime,t) d\vec{\chi}$ and $M_{ij}^2 = M_{ij}^2(x_i^\prime,t,\tau) = \int \Delta_i \Delta_j P(\vec{\chi},t+\tau|\vec{x}^\prime,t) d\vec{\chi}$ are the first and second moments, respectively. It is assumed that higher-order moments are zero \cite{risken1996fokker}. 

To obtain the moments, we first write the evolution of \eq{eq:dynamics} in small time interval $\tau$
\begin{equation}
\label{eq:moments_1}
\vec{x}^\prime(t+\tau) = \vec{x}^\prime(t) + \vec{v}(\vec{x}^\prime(t),t)\tau + \mat{G}(\vec{x}^\prime(t),t)d\vec{W}(t) ,
\end{equation}
where $d\vec{W} = d\vec{\eta}(t)\tau$. The delta function $\delta(\vec{x}^\prime - \vec{x})$ in \eq{eq:FP_KM} implies that the dynamics is valid at $\vec{x}^\prime$ close to $\vec{x}$.  If there exists a condition in such a way that $\vec{x}^\prime - \vec{x} = \vec{v}_f\tau$ where $\vec{v}_f$ is arbitrary finite velocity, we can write $\vec{v}(\vec{x},t) \equiv \vec{v}(\vec{x}^\prime,t) - \frac{\partial \vec{v}}{\partial \vec{x}} \vec{v}_f\tau + \Order{\tau^2}$. 
Gathering all terms, the first moment can be calculated from this formula \cite{risken1996fokker}
\begin{eqnarray}
\label{eq:M1}
M_i^1 &=& \langle \chi_{i}(t+\tau)-\chi_{i}(t)-v_{i}(x_i^\prime,t)\tau + \Order{\tau^2} \rangle \vert_{\chi_i=x_i^\prime}  \nonumber\\
&=& \langle G_{im} dW_m \rangle = 0,
\end{eqnarray}
where $\vert_{\chi_i=x_i^\prime}$ means that the stochastic variable $\chi_i$ at time $t$ has the sharp value $x_i^\prime$ \cite{risken1996fokker}.
The second moment can be calculated from this formula \cite{risken1996fokker}
\begin{eqnarray}
\label{eq:M2}
M_{ij}^2 &=& \langle (x_{i}^\prime(t+\tau)-x_{i}^\prime(t)-v_{i}(x_i^\prime,t)\tau + \Order{\tau^2}) \nonumber\\
&& \times (x_{j}^\prime(t+\tau)-x_{j}^\prime(t)-v_{j}(x_j^\prime,t)\tau + \Order{\tau^2}) \rangle \nonumber\\
&=& \langle G_{im} G_{jn} dW_m dW_n \rangle = G_{im} G_{jn}Q_{mn} \tau .
\end{eqnarray}
Substituting \eq{eq:M1} and \eq{eq:M2} into \eq{eq:FP_KM}, we have
\begin{equation}
\label{eq:FP_KM_2}
P(\vec{x}+\vec{v}\tau,t+\tau|\vec{x}^\prime,t) 
=  \left(1 + \tau D_{ij}^\prime \frac{\partial^2 }{\partial x_{i}^\prime \partial x_{j}^\prime }  \right) \delta(\vec{x}^\prime - \vec{x}),
\end{equation}
where $D_{ij}^\prime = \frac{1}{2} M_{ij}^2(x_i^\prime,t,\tau)$. Using \eq{eq:FP_int_sol} and \eq{eq:FP_KM_2}, we have
\begin{eqnarray}
\label{eq:FP_int_sol_2}
p(\vec{x}+\vec{v}\tau,t+\tau) &=& \int P(\vec{x}+\vec{v}\tau,t+\tau|\vec{x}^\prime,t)  p(\vec{x}^\prime,t) d\vec{x}^\prime \nonumber\\
&=& \left( 1 + \tau \frac{\partial^2 D_{ij} }{\partial x_i\partial x_j} \right) p(\vec{x},t),
\end{eqnarray}
where we use the property of the delta function in such a way that $\int dx f(x) \frac{\partial^k }{\partial x^k} \delta(x-y) = (-1)^k \frac{\partial^k }{\partial x^k}f(x) \vert_{x=y} $ \cite{apaza2020homotopy}. When we expand \eq{eq:FP_int_sol_2} further in the term of a small time interval $\tau$, we have the following
\begin{equation}
\label{eq:FP_new}
\frac{D p}{D t} = \frac{\partial p}{\partial t} + \sum_{i}^{n} v_i\frac{\partial p}{\partial x_i} + \Order{\tau} =  \frac{\partial^2 D_{ij} }{\partial x_i\partial x_j} p ,
\end{equation}
where $\frac{D (\cdot)}{D t} \equiv \frac{\partial (\cdot)}{\partial t} + \sum_{i}^{n} v_i\frac{\partial (\cdot)}{\partial x_i}$ is \emph{material derivative} or total derivative and $p \equiv p(\vec{x},t|\vec{x}^\prime,t^\prime)$.  It should be noted that the dynamics, as in \eq{eq:dynamics}, have a vanishing divergence $\sum_{i}^{n} \frac{\partial v_i }{\partial x_i} = 0$. With this property, \eq{eq:FP_new} recovers the correct Fokker-Planck equation as in \eq{eq:FP}. The advantage of \eq{eq:FP_new} is the fact that the non-linear drift function can be hidden and the system remains only a diffusion process. This equation is an alteration of the Fokker-Plack equation which can be used to examine a range of stochastic processes that are represented by \eq{eq:dynamics} with a vanishing divergence $\sum_{i}^{n} \frac{\partial v_i }{\partial x_i} = 0$.

\subsection{Short-time drift propagator}
For small time differences $\tau=t-t^\prime \ll 1$, \eq{eq:FP_KM_2} may be represented in the following way by using the procedure in the standard textbook \cite{risken1996fokker}
\begin{equation}
\label{eq:P_short}
P(\vec{x}+\vec{v}\tau,t+\tau|\vec{x}^\prime,t) = \left[ 1 + \tau\Opt{L}  + \Order{\tau^2} \right]\delta(\vec{x}-\vec{x}^\prime) ,
\end{equation}
where $\Opt{L}[(\cdot)] \equiv  D_{ij} \frac{\partial^2 (\cdot) }{\partial x_i\partial x_j}$. The difference between our approach to the standard one is that the system transits from $\vec{x}^\prime$ to $\vec{x}+\vec{v}\tau$ instead of $\vec{x}$, during a short-time interval $\tau=t-t^\prime \ll 1$. Thus $P(\vec{x}+\vec{v}\tau,t+\tau|\vec{x}^\prime,t)$ is called a short-time drift propagator. 

Up to $\Order{\tau^2}$, \eq{eq:P_short} can be approximate
\begin{equation}
\label{eq:P_short_exp}
P(\vec{x}+\vec{v}\tau,t+\tau|\vec{x}^\prime,t) \approx  e^{ \left( \tau D_{ij} \frac{\partial^2 }{\partial x_i\partial x_j}  \right) } \delta(\vec{x}-\vec{x}^\prime).
\end{equation}
Substituting the expression for the delta function of N variables $\delta(\vec{x}-\vec{x}^\prime) = (2\pi)^{-N}\int\exp[iu_j(x_j-x_j^\prime)]d\vec{u}$ (where $i=\sqrt{-1}$) into \eq{eq:P_short_exp} and doing some calculation \cite{risken1996fokker}, we obtain a short-time drift propagator
\begin{eqnarray}
\label{eq:dift_P_1}
\lefteqn{
P(\vec{x}+\vec{v}\tau,t+\tau|\vec{x}^\prime,t) } \nonumber\\
&=& \frac{1}{(2\pi)^N} e^{ \left( \tau D_{ij} \frac{\partial^2 }{\partial x_i\partial x_j} \right) } \int\exp[iu_j(x_j-x_j^\prime)] d\vec{u} \nonumber\\
&=& \frac{1}{(2\pi)^N} \int e^{ \left[ -\tau D_{jk} u_j u_k +  iu_j(x_j-x_j^\prime)  \right] } d\vec{u} \nonumber\\
&=& \frac{1}{(2\pi)^N} \int e^{ \left[ -\tau \vec{u}^\top \mat{D} \vec{u} +  i\vec{u}^\top (\vec{x} - \vec{x}^\prime)  \right] } d\vec{u}.
\end{eqnarray}
The covariance matrix can be factored such that 
\begin{equation}
    \mat{D} = \mat{S}\mat{S}^\top 
\end{equation}
where $\mat{S}^{-1}\mat{D}(\mat{S}^{-1})^\top = \mat{I}$ \cite{stark1986probability}. We change the variables 
\begin{eqnarray}
    \vec{w} &=& \sqrt{2\tau} \mat{S}^\top\vec{u} - \vec{\mu} \\
    \vec{\mu} &=& i\frac{\mat{S}^{-1}}{\sqrt{2\tau}} (\vec{x} - \vec{x}^\prime).
\end{eqnarray}
With these transformations, the exponent on the right-hand side of \eq{eq:dift_P_1} becomes 
\begin{eqnarray}
\lefteqn{
    -\tau \vec{u}^\top \mat{D} \vec{u} +  i\vec{u}^\top (\vec{x} - \vec{x}^\prime)
} \nonumber \\
    && = -\frac{1}{2} (\vec{w}+\vec{\mu})^\top (\vec{w}+\vec{\mu}) + (\vec{w}+\vec{\mu})^\top \vec{\mu} \nonumber \\
    && = -\frac{1}{2} \vec{w}^\top \vec{w} + \frac{1}{2} \vec{\mu}^\top \vec{\mu}.
\end{eqnarray}
For a linear transformation, the volume elements are related as $d\vec{w} = \sqrt{2\tau}^N |\mat{S}^\top| d\vec{u}$. Gathering all terms, \eq{eq:dift_P_1} becomes
\begin{eqnarray}
\label{eq:dift_P_2}
\lefteqn{
P(\vec{x}+\vec{v}\tau,t+\tau|\vec{x}^\prime,t) } \nonumber\\
&=& \frac{ e^{\frac{1}{2} \vec{\mu}^\top \vec{\mu}} }{ \sqrt{2\pi}^{N} }  \int \frac{ e^{-\frac{1}{2} \vec{w}^\top \vec{w}} }{ \sqrt{2\pi}^{N} } \frac{ |(\mat{S}^{-1})^\top| }{ \sqrt{2\tau}^{N} } d\vec{w} .
\end{eqnarray}
The term $\int \sqrt{2\pi}^{-N} e^{-\frac{1}{2} \vec{w}^\top \vec{w}} d\vec{w}$ is known as multivariate Gaussian integral and equal to unity. Since $|\mat{D}^{-1}| = |\mat{S}^{-1}(\mat{S}^{-1})^\top| = |(\mat{S}^{-1})^\top|^2$, it shows that $|(\mat{S}^{-1})^\top| = \sqrt{|\mat{D}^{-1}|}$. Finally, after calculating, we obtain the transition probability function 
\begin{equation}
\label{eq:diftP}
P(\vec{x}+\vec{v}\tau,t+\tau|\vec{x}^\prime,t) = \frac{ e^{-\frac{1}{4\tau} (\vec{x}-\vec{x}^\prime)^\top \mat{D}^{-1} (\vec{x}-\vec{x}^\prime)} }{\sqrt{(4\pi\tau)^N |\mat{D}|}} .
\end{equation}
With \eq{eq:FP_int_sol} and \eq{eq:diftP}, the short-time solution of \eq{eq:FP} can be found
\begin{equation}
\label{eq:FP_int_sol_short}
p(\vec{x}+\vec{v}\tau,t+\tau) = \int \frac{ e^{-\frac{1}{4\tau} (\vec{x}-\vec{x}^\prime)^\top \mat{D}^{-1} (\vec{x}-\vec{x}^\prime)} }{\sqrt{(4\pi\tau)^N |\mat{D}|}} p(\vec{x}^\prime,t) d\vec{x}^\prime .
\end{equation}

This novel approach in \eq{eq:FP_int_sol_short} is different from traditional solutions \eq{eq:FP_int_sol} and \eq{eq:propagator_old}, as the state dependence function $\vec{v}$ has been removed from the integral term and instead appears on the left-hand side as a state-shifted term. The right-hand side of \eq{eq:FP_int_sol_short} is a straightforward multivariate Gaussian integral, which we will discuss in the following section. Implementing the state change term $\vec{x}+\vec{v}\tau$ is difficult due to its dependence on the intricate dynamics of the system. When $\vec{v}$ is constant, the state space can be represented as a regular grid, making it easy to perform the calculation, as we will see in the numerical examples section.

\subsection{Gaussian-Hermit quadrature solution}
\eq{eq:FP_int_sol_short} only contains a multidimensional Gaussian integral that can be numerically solved by the Gaussian-Hermit quadrature equation \cite{abramowitz1972handbook, NumericalRecipes}.

We change the variable $\vec{\varepsilon} = \frac{1}{\sqrt{4\tau}} \mat{S}^{-1}(\vec{x}-\vec{x}^\prime)$, where $\mat{D}=\mat{S}\mat{S}^\top$. Applying this into \eq{eq:FP_int_sol_short}, we have
\begin{equation}\label{eq:FP_Gaussian}
    p(\vec{x}+\vec{v}\tau,t+\tau) =  \frac{1}{ \sqrt{\pi^N} } \int  e^{ -\vec{\varepsilon}^\top \vec{\varepsilon}} p(\vec{x}-\sqrt{4\tau}\mat{S}\vec{\varepsilon},t) d\vec{\varepsilon} . 
\end{equation}
Since the exponential term in \eq{eq:FP_Gaussian} can be written as $e^{-\sum_i \varepsilon_i^2}$, the integral can now be split into nested Gauss-Hermite integrals \cite{jackel2005note}
\begin{eqnarray}\label{eq:FP_GH}
    \lefteqn{
    p(\vec{x}+\vec{v}\tau,t+\tau) = } \nonumber \\
    &&  \frac{1}{ \sqrt{\pi^N} } \sum_{i_1=1}^{m_1} \sum_{i_2=1}^{m_2}  \cdots \sum_{i_N=1}^{m_N} w_{i_1} w_{i_2} \cdots w_{i_N} \nonumber \\
    && \times p(a_1, a_2, \dots, a_N, t) ,
\end{eqnarray}
where $a_j = x_j-\sqrt{4\tau}S_j \varepsilon_{i_j}$, $\varepsilon_{i_j}$ is root of the Hermite polynomials of order $m_j$ and $w_{i_j}$ is associated weight. 

\section{Numerical Examples}
In this section, we will demonstrate the use of our numerical algorithm by examining a simple example of a linear system of equations. We will consider a basic drift-diffusion problem that is described by a Wiener process \cite{risken1996fokker, frank2005nonlinear}.

\subsection{1D drift-diffusion process}
\begin{figure}
     \centering
     \begin{subfigure}[b]{0.4\textwidth}
         \centering
         \includegraphics[width=\textwidth]{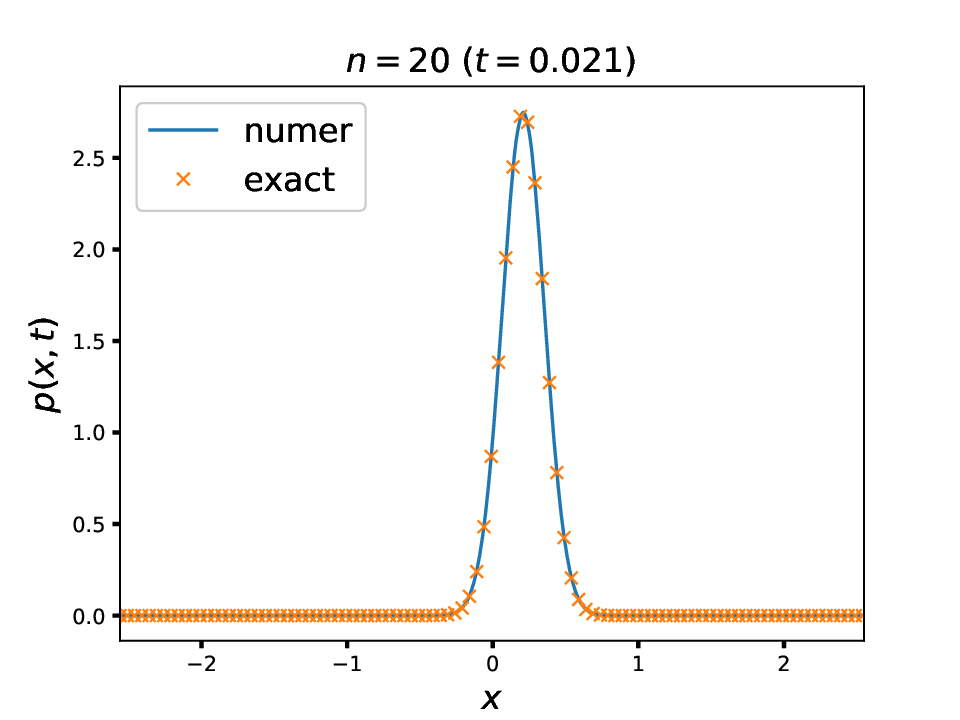}
     \end{subfigure}
     \hfill
     \begin{subfigure}[b]{0.4\textwidth}
         \centering
         \includegraphics[width=\textwidth]{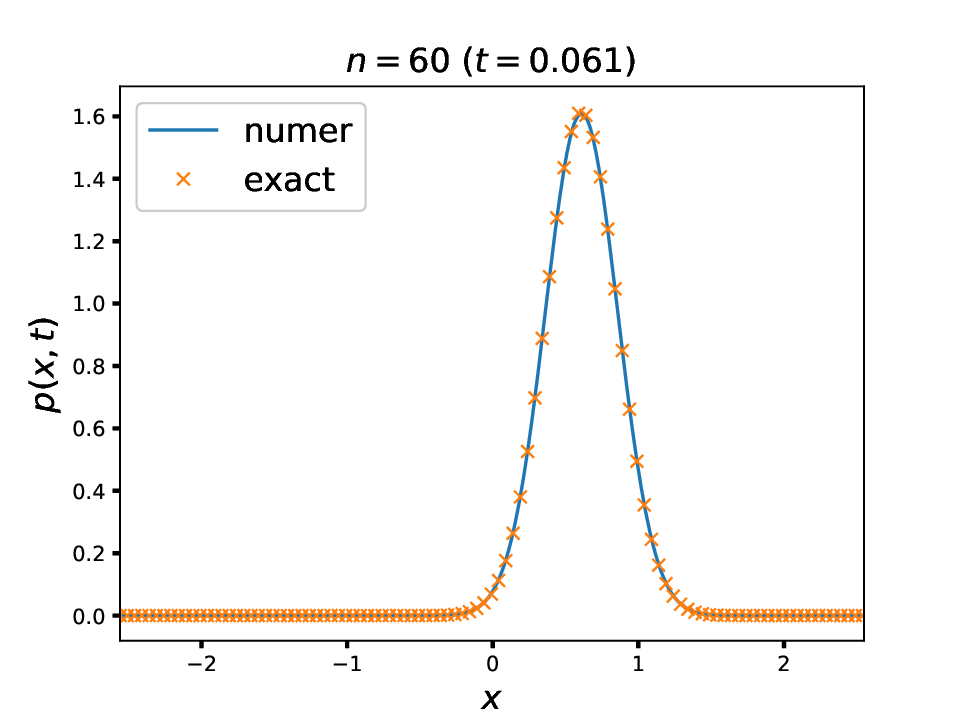}
     \end{subfigure}
     \hfill
     \begin{subfigure}[b]{0.4\textwidth}
         \centering
         \includegraphics[width=\textwidth]{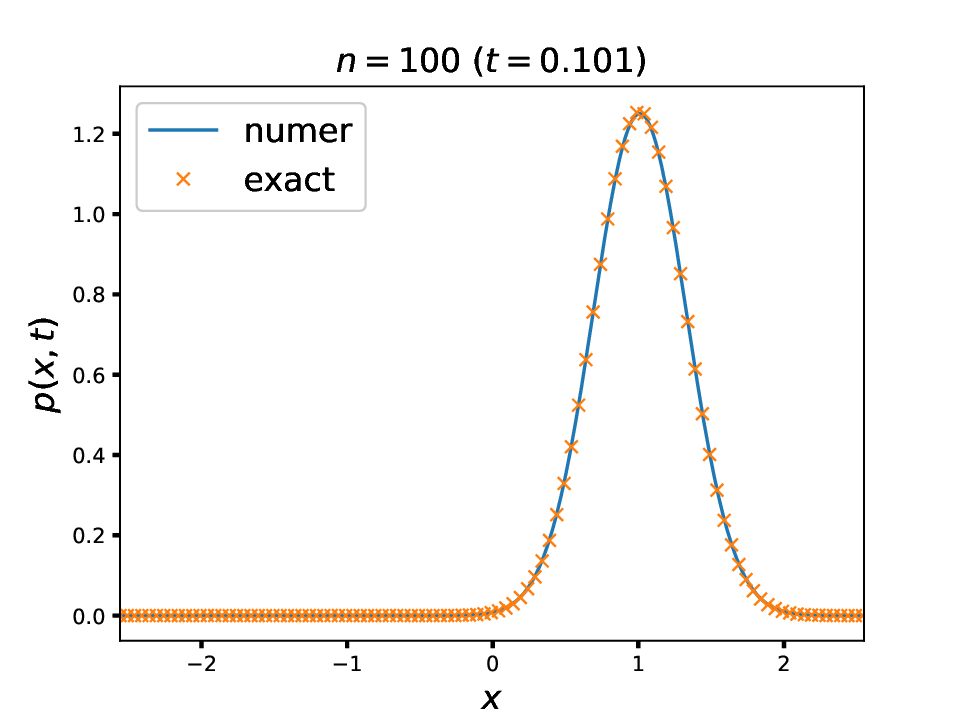}
     \end{subfigure}
        \caption{
A comparison of the numerical probability density profile to the exact solution in some selected time step for 1D drift-diffusion system.
        }
        \label{fig:p1d}
\end{figure}

We initially investigated the drift-diffusion process in a one-dimensional space (1D). The Wiener equation is expressed as follow:
\begin{equation}\label{eq:Wiener_1D}
    \dot{x}(t) = v + \sigma\eta(t),
\end{equation}
where $v$ is constant drift velocity and $\sigma$ is constant noise strength (standard deviation). The Fokker-Planck equation for \eq{eq:Wiener_1D} is expressed as follow: 
\begin{equation}\label{eq:FP_1d_drift_diffusion}
    \frac{\partial }{\partial t} p(x,t) + v\frac{\partial}{\partial x}p(x,t) = \frac{\sigma^2}{2}\frac{\partial^2}{\partial x^2}p(x,t) .
\end{equation} 
The solution to drift-diffusion equation \eq{eq:FP_1d_drift_diffusion} with delta function initial condition $p(x,0)=\delta(x)$ is expressed in the form of a Gaussian function \cite{risken1996fokker, frank2005nonlinear}:
\begin{equation}\label{eq:1d_drift_diffusion_sol}
    p(x,t) = \frac{1}{\sqrt{2\pi \sigma^2 t}} e^{- \frac{1}{2} \frac{(x-vt)^2}{\sigma^2 t}} .
\end{equation}

We will employ numerical techniques to address the 1D drift-diffusion equation with the algorithms proposed in \eq{eq:FP_GH}. To program \eq{eq:FP_GH}, the space and time variables must be changed into discrete values, which are specified as follows: $x_j = jh$ and $t_n = n\tau$, where $h=v\tau$ is the spacing step, $j = \{0, 1, 2, \dots, j_\textnormal{max}\}$ and $n = \{0, 1, 2, \dots, n_\textnormal{max}\}$. We select a 3-point Gauss-Hermite quadrature rule, for which the roots and associated weights are given by the following values: $\varepsilon_\alpha = \{-1.22474487, 0, 1.22474487\}$ and $w_\alpha = \{0.29540898, 1.1816359, 0.29540898\}$ \cite{abramowitz1972handbook}. In the 1D system, the formula in \eq{eq:FP_GH} with the 3-point Gauss-Hermite quadrature rule can be expressed as
\begin{equation}\label{eq:Algorithm_1d}
    p(x_j + h, t_n + \tau) = \frac{1}{\sqrt\pi}\sum_{i=1}^3 w_\alpha p(x_j + \xi_\alpha, t_n) ,
\end{equation}
where $\xi_\alpha = \varepsilon_\alpha \sigma\sqrt{2\tau}$. The value of the distribution function $p$ at an arbitrary position $x_j + \varepsilon_\alpha \sigma\sqrt{2\tau}$ on the right side of \eq{eq:Algorithm_1d} can be determined by linear interpolation \cite{NumericalRecipes}. Algorithm \eqa{eq:Algorithm_1d} can be divided into three stages for each iteration:
\begin{algorithm}
(1) $\hat{p}(x, t_n) \gets \textnormal{ linear interpolate} ~p(x_j, t_n)$ \\
(2) $p^\ast(x_j, t_n) \gets \frac{1}{\sqrt\pi}\sum_{i=1}^3 w_\alpha \hat{p}(x_j + \xi_\alpha, t_n)$ \\
(3) $p(x_j + h, t_n + \tau) \gets p^\ast(x_j, t_n)$ . \\

\caption{Numerical scheme for a 1D drift-diffusion system}
\end{algorithm}

We perform our numerical calculations on 512 grid points with the parameters $h=0.01$, $\tau=0.001$, and $\sigma=1$, resulting in $c=10$. The initial distribution is set to the exact solution given by \eq{eq:1d_drift_diffusion_sol} with the initial time slightly greater than zero, $t_0 = 0.001$, in order to avoid an infinite value. We employ the periodic boundary condition for \eq{eq:Algorithm_1d}, making $p(x_0,t_n)$ equivalent to $p(x_{j_\textnormal{max}}, t_n)$. 

Figure \ref{fig:p1d} displays a comparison between the numerical probability density profile ($p_\textnormal{numer}$) and the exact solution ($p_\textnormal{exact}$) in a chosen time step for a 1D drift-diffusion system. It has been observed that the probability density profile moves to the right as expected. The numerical results and the exact solution are in agreement, with an average absolute error of approximately $\sim 10^{-4}$, where the absolute error is determined by $\mid p_\textnormal{numer} - p_\textnormal{exact} \mid$.

\subsection{2D drift-diffusion process}
\begin{figure}
     \centering
     \begin{subfigure}[b]{0.4\textwidth}
         \centering
         \includegraphics[width=\textwidth]{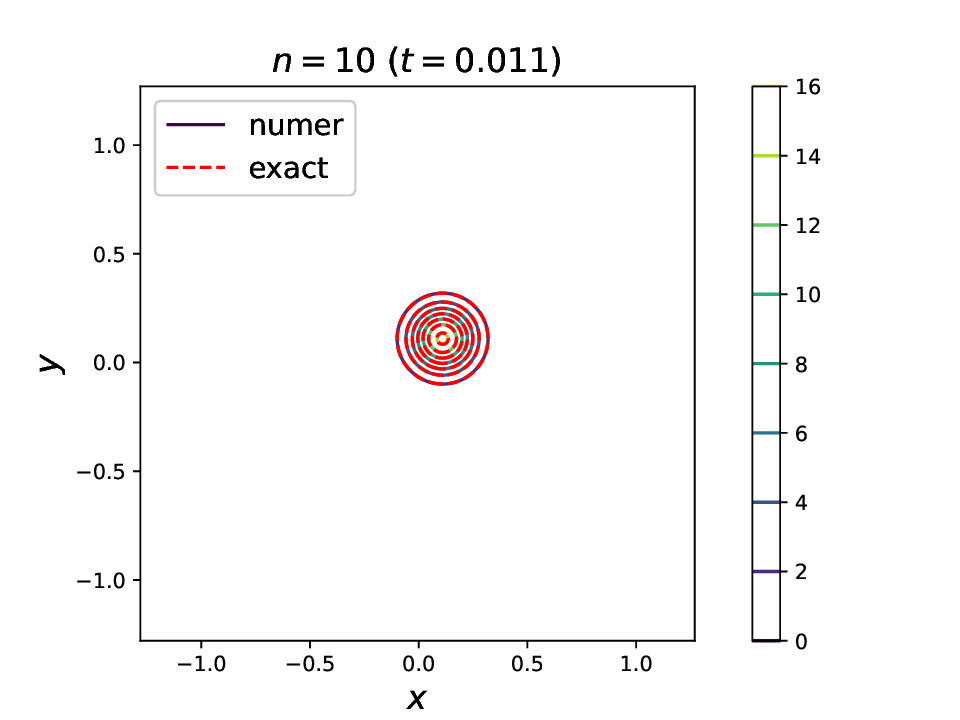}
     \end{subfigure}
     \hfill
     \begin{subfigure}[b]{0.4\textwidth}
         \centering
         \includegraphics[width=\textwidth]{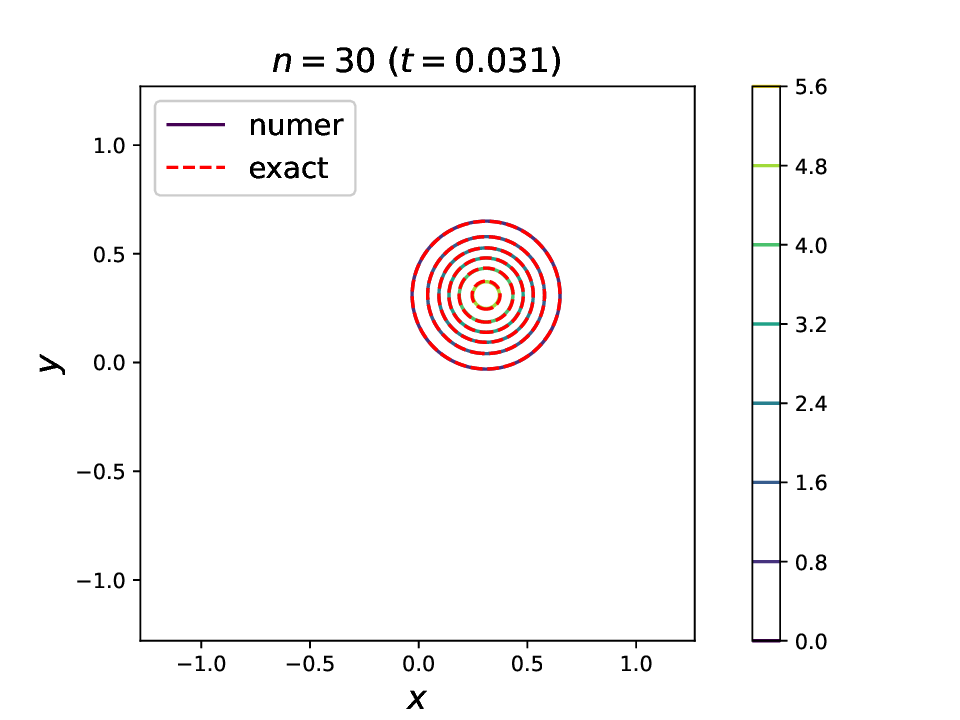}
     \end{subfigure}
     \hfill
     \begin{subfigure}[b]{0.4\textwidth}
         \centering
         \includegraphics[width=\textwidth]{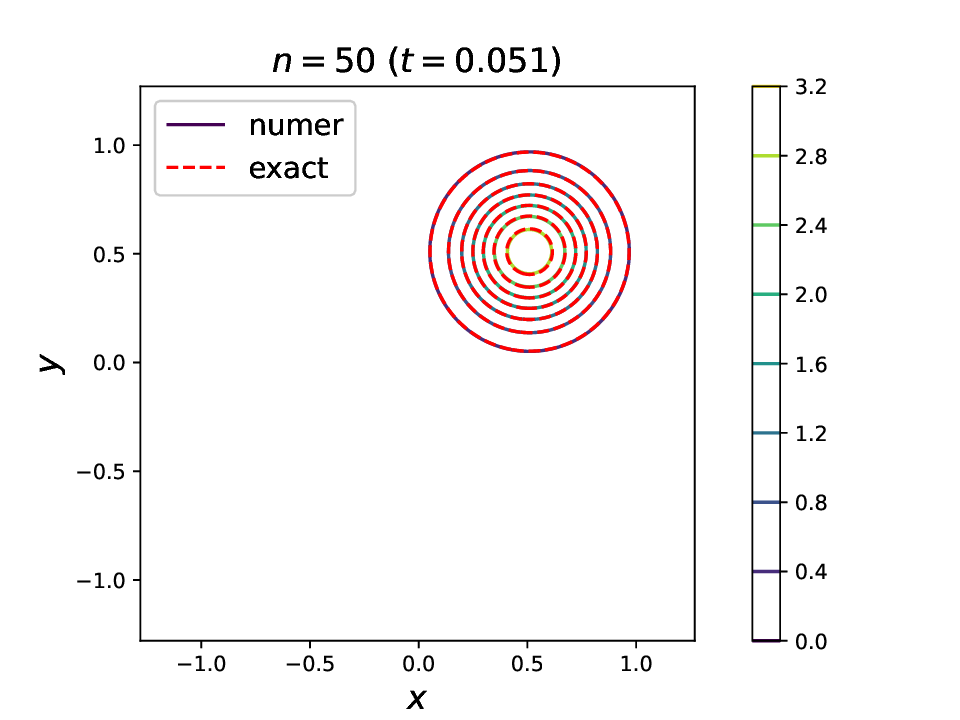}
     \end{subfigure}
        \caption{
A comparison of the numerical probability density profile to the exact solution in some selected time step for 2D drift-diffusion system.
        }
        \label{fig:p2d}
\end{figure}
We extend our method to investigate the isotropic drift-diffusion process in two-dimensional space (2D). The Wiener equation is expressed as:
\begin{equation}\label{eq:Wiener_2D}
    \underbrace{\begin{bmatrix}
    \dot{x}(t)\\ 
    \dot{y}(t)
    \end{bmatrix}}_{\dot{\vec{x}}(t)} =
    \underbrace{\begin{bmatrix}
    v_x \\ 
    v_y
    \end{bmatrix}}_{\vec{v}}  + 
    ~\sigma\underbrace{\begin{bmatrix}
    \eta_x(t)\\ 
    \eta_y(t)
    \end{bmatrix}}_{\vec{\eta}(t)} ,
\end{equation}
where the drift velocity in the $x$ and $y$ directions is denoted by $v_x$ and $v_y$ respectively, and the noise strength (standard deviation) in both directions is the same and is represented by $\sigma$. The drift velocity and noise strength remain constant. The Fokker-Planck equation for \eq{eq:Wiener_2D} is expressed as follow: 
\begin{equation}\label{eq:FP_2d_drift_diffusion}
    \frac{\partial }{\partial t} p(\vec{x},t) + \vec{v}\cdot\nabla p(\vec{x},t) = \frac{\sigma^2}{2}\nabla^2 p(\vec{x},t) .
\end{equation} 
The solution to drift-diffusion equation \eq{eq:FP_2d_drift_diffusion} with delta function initial condition $p(x,y,0) = \delta(x)\delta(y)$ is expressed in the form of a Gaussian function \cite{risken1996fokker, frank2005nonlinear}:
\begin{equation}\label{eq:2d_drift_diffusion_sol}
    p(x,y,t) = \frac{1}{2\pi \sigma^2 t} e^{- \frac{1}{2} \frac{(x-v_xt)^2 + (y-v_yt)^2}{\sigma^2 t}} .
\end{equation}

We will numerically solve the 2D drift-diffusion equation using the algorithms we have proposed in equation \eq{eq:FP_GH}. To program \eq{eq:FP_GH}, the space and time variables must be changed into discrete values, which are specified as follows: $x_j = jh_x$, $y_k = kh_y$ and $t_n = n\tau$, where $h_x=v_x\tau$ and $h_y=v_y\tau$ are the spacing steps in $x$ and $y$ directions, respectively, $j = \{0, 1, 2, \dots, j_\textnormal{max}\}$, $k = \{0, 1, 2, \dots, k_\textnormal{max}\}$ and $n = \{0, 1, 2, \dots, n_\textnormal{max}\}$. We select a 3-point Gauss-Hermite quadrature rule, for which the roots and associated weights are given by the following values: $\varepsilon_{\alpha (\beta)} = \{-1.22474487, 0, 1.22474487\}$ and $w_{\alpha (\beta)} = \{0.29540898, 1.1816359, 0.29540898\}$ \cite{abramowitz1972handbook}. In the 2D system, the formula in \eq{eq:FP_GH} with the 3-point Gauss-Hermite quadrature rule can be expressed as
\begin{eqnarray}\label{eq:Algorithm_2d}
\lefteqn{
    p(x_j + h_x, y_k + h_y, t_n + \tau) = 
} \nonumber\\
    &&  \frac{1}{\pi}\sum_{\alpha=1}^3 \sum_{\beta=1}^3 w_{\alpha} w_{\beta} p(x_j + \xi_\alpha, y_k + \xi_\beta, t_n) ,
\end{eqnarray}
where $\xi_{\alpha (\beta)} = \varepsilon_{\alpha (\beta)} \sigma\sqrt{2\tau}$. The value of the distribution function $p$ at an arbitrary position $x_j + \varepsilon_\alpha \sigma\sqrt{2\tau}$ and $y_k + \varepsilon_\beta \sigma\sqrt{2\tau}$  on the right side of \eq{eq:Algorithm_2d} can be determined using bilinear interpolation \cite{NumericalRecipes}. Algorithm \eqa{eq:Algorithm_2d} can be divided into three parts for each iteration:

\begin{algorithm}
(1) $\hat{p}(x, y, t_n) \gets \textnormal{bilinear interpolate} ~p(x_j, y_k, t_n)$ \\
(2) $p^\ast(x_j, y_k, t_n) \gets \frac{1}{\pi}\sum_{\alpha=1}^3 \sum_{\beta=1}^3 w_{\alpha} w_{\beta} \hat{p}(x_j + \xi_\alpha, y_k + \xi_\beta, t_n)$ \\
(3) $p(x_j + h_x, y_k + h_y, t_n + \tau) \gets p^\ast(x_j, y_k, t_n)$ \\

\caption{Numerical scheme for a 2D drift-diffusion system}
\end{algorithm}

We perform our numerical calculations on a $256\times 256$ grid point with the parameters $h_x=h_y=0.01$, $\tau=0.001$, and $\sigma=1$, which results in $c_x=c_y=10$. The initial distribution is set to the exact solution given by \eq{eq:2d_drift_diffusion_sol} with the initial time slightly greater than zero, $t_0 = 0.001$, in order to avoid an infinite value. We employ the periodic boundary condition for \eq{eq:Algorithm_2d}, making $p(x_0,t_n)$ equivalent to $p(x_{j_\textnormal{max}}, t_n)$ and $p(y_0,t_n)$ equivalent to $p(y_{k_\textnormal{max}}, t_n)$. 

The contour plot in Figure \ref{fig:p2d} shows the numerical probability density ($p_\textnormal{numer}$) compared to the exact solution ($p_\textnormal{exact}$) at a certain time step for a two-dimensional drift-diffusion system. It has been seen that the probability density profile shifts in the direction of $\vec{v}$ as expected. Again, as in 1D, the numerical results and the exact solution agree, with an average absolute error of approximately $\sim 10^{-3}$, where the absolute error is determined by $\mid p_\textnormal{numer} - p_\textnormal{exact} \mid$.

\section{Conclusion}
In summary, we propose a new approach to the solution of the Fokker-Planck equation within short-time intervals. The numerical solution of the Fokker-Planck equation can be obtained iteratively using a new technique based on a short-time drift propagator. We believe that it will be of great interest to researchers from many fields.

\begin{acknowledgments}
This research project was supported by the Thailand Science Research and Innovation Fund and the University of Phayao (\emph{Grant no. FF65-RIM070}).
\end{acknowledgments}


\bibliography{ref}

\end{document}